%% file: main.tex
\documentclass[sigconf,nonacm]{acmart}

\newtheorem{theorem}{\textbf{Theorem}}
\newtheorem{definition}{\textbf{Definition}} 

\newtheorem{property}{\textbf{Property}}
\newtheorem{lemma}[theorem]{\textbf{Lemma}}

\usepackage{amsmath}
\usepackage{tabularx}
\usepackage{booktabs,multirow,graphicx}
\usepackage{algpseudocode}
\usepackage[lined,linesnumbered,boxed,commentsnumbered, ruled,vlined]{algorithm2e}

\usepackage{cuted}
\usepackage{graphicx}
\usepackage{xspace}
\usepackage{soul}
\usepackage{tikz}
\usepackage[normalem]{ulem}
\usepackage{enumitem}
\usetikzlibrary{arrows.meta,positioning,fit}
\SetKw{Continue}{continue}

\newcommand{\ie}{\emph{i.e.,}\xspace}

\definecolor{revisionred}{RGB}{220,0,0}

\AtBeginDocument{%
  }

\begin{document}

\title{A Gaussian Primitive Data Management Framework for Efficient 3DGS Trajectory Rendering}
\author{Haolun Ma}
\affiliation{%
  \institution{DSA Thrust, HKUST(GZ), China}
}
\email{hma987@connect.hkust-gz.edu.cn}

\author{Ziyi Liu}
\affiliation{%
  \institution{Soochow University, China}
}
\email{liuziyi@suda.edu.cn}

\author{Zizhuo Xu}
\affiliation{%
  \institution{HKUST, Hong Kong SAR, China}
}
\email{zxucj@connect.ust.hk}

\author{Chuhan Hu}
\affiliation{%
  \institution{DSA Thrust, HKUST(GZ), China}
}
\email{chu083@connect.hkust-gz.edu.cn}

\author{Lei Li}
\affiliation{%
  \institution{DSA Thrust, HKUST(GZ), China}
}
\email{thorli@ust.hk}

\begin{abstract}
Trajectory rendering, which renders novel views of a reconstructed 3D scene along a camera path, is a core operation in digital twins and autonomous inspection. Structured 3D Gaussian Splatting (3DGS) renders a single view in real time by decoding sparse anchors into Gaussians on demand, but it renders a trajectory view by view: every view projects all anchors to select them, re-decodes anchors already requested by neighboring views, and runs all steps on the GPU while the CPU idles. We treat this as a spatial data management problem, where anchors are spatial objects, views are range queries, and decoded Gaussians are shared derived data, and propose GDM-GS, which removes redundancy at three levels. Within a view, an anchor index and an occluder index answer each view as a holed-frustum range query, so out-of-view and provably hidden anchors are never decoded. Across views, shared materialization decodes each requested anchor once per group of consecutive views while preserving each view's selection. Across processors, a hybrid pipeline balances selections between the CPU and the GPU before grouped decoding under bounded memory. Experiments on large-scale public scenes show that GDM-GS substantially reduces trajectory rendering latency while preserving image quality.

\end{abstract}

\maketitle

\pagestyle{plain}

\input{sections/introduction}

\input{sections/preliminary}

\input{sections/Overview}

\input{sections/method}

\input{sections/experiments}
\input{sections/relatedwork}
\input{sections/conclusion}

\balance
\bibliographystyle{ACM-Reference-Format}
\bibliography{references}

\end{document}

%% file: sections/introduction.tex
\section{Introduction}
\label{sec:intro}

\begin{figure}
    \centering
    \includegraphics[width=1\linewidth]{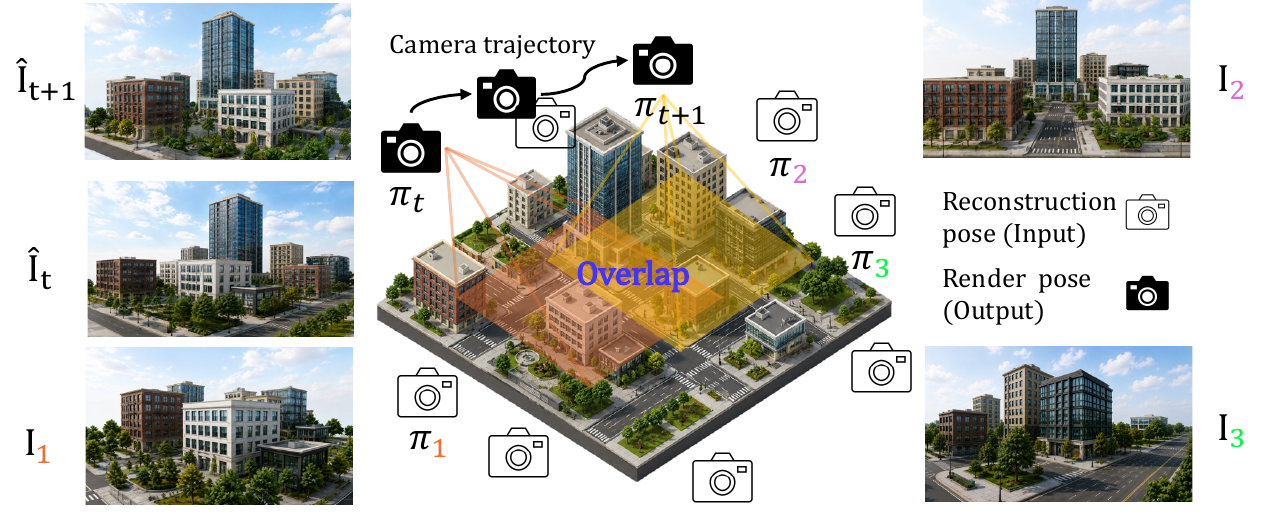}
    \vspace*{-4mm}
    \caption{Multi-View Reconstruction \& Trajectory Rendering}
    \label{fig:nvs}
    \vspace*{-8mm}
\end{figure}

\textit{Novel View Synthesis (NVS)}~\cite{mildenhall2020nerf,kerbl2023gs} reconstructs a 3D scene from multi-view images with calibrated camera poses and renders photorealistic images from unseen viewpoints. It has been applied to digital-twin modeling \cite{guo2025articulatedgs}, immersive virtual environments \cite{franke2025vrsplatting}, and visual inspection \cite{trombini2026inspection}. Most existing studies focus on \textit{single-pose NVS}, where each target view is synthesized independently, \ie rendering a digital-twin scene from a user-specified viewpoint.

As NVS is increasingly adopted by mobile and autonomous applications, such as autonomous driving~\cite{yang2025drivex}, low-altitude aerial inspection~\cite{zaouali2025flight}, and mobile robotics~\cite{chhablani2025embodied}, the rendering demand is shifting from isolated viewpoints to continuous camera trajectories. These applications require sequences of novel views along the motion path to support navigation, inspection, and scene understanding. We refer to this setting as \textit{trajectory-based NVS}, where a sequence of novel views is rendered along a predefined ordered camera trajectory over a fixed reconstructed scene. As shown in Figure~\ref{fig:nvs}, a scene is first reconstructed offline from multi-view image--pose pairs, illustrated by $(I_1,\pi_1)$, $(I_2,\pi_2)$, and $(I_3,\pi_3)$, and is then rendered online along a camera trajectory, where consecutive poses $\pi_t$ and $\pi_{t+1}$ produce consecutive frames $\hat{I}_t$ and $\hat{I}_{t+1}$. When the trajectory is smooth at frame granularity, the fields of view of these poses overlap substantially, so frame-by-frame rendering repeats a large amount of computation over the same underlying scene data.

Existing NVS methods mainly on two scene representations: 1) \textit{Neural Radiance Fields (NeRF)}~\cite{mildenhall2020nerf} represent a scene as an implicit neural field and render views through dense ray sampling and neural evaluation. Due to the high computational cost of high-resolution synthesis, NeRF is not considered. 2) \textit{3D Gaussian Splatting (3DGS)}~\cite{kerbl2023gs} represents a scene using explicit spatial Gaussian primitives that can be efficiently projected and rendered. Structured methods such as Scaffold-GS and Octree-GS~\cite{lin2024scaffoldgs,ren2024octree} replace explicit storage of all Gaussians with sparse anchors and generate Gaussians on demand through a shared decoder. 

Specifically, rendering a view takes three steps: \textit{Selecting candidate anchors} (S1), \textit{Materializing Gaussians} (S2), and \textit{Rendering the image} (S3). Here, S1 acts as a spatial query over the anchors, S2 materializes the selected anchors into Gaussians, and S3 consumes these Gaussians to produce the image. Existing methods tightly couple these steps in a per-view GPU pipeline, leaving little opportunity to avoid reusing results across views. This overhead becomes substantial over a trajectory of $T$ views and $N_a$ anchors, leading to the following sources of waste:

\begin{enumerate}
    \item \textbf{\textit{Full-scan Selection}:}
    S1 scans a set of anchors for every view and evaluates eligible anchors through individual projection tests. Selection therefore incurs $O(T\cdot N_a)$ scan work over the trajectory, even when each view covers only a small part of the scene. Moreover, passing an anchor-level test does not ensure that its Gaussians contribute to the image: they may still be rejected by finer tests or hidden behind nearer surfaces. Such Gaussians incur decoding and rendering preparation before their lack of contribution is determined.
    \item \textbf{\textit{Repeated Materialization:}}
    Consecutive views along a smooth trajectory often select many of the same anchors. However, S2 independently decodes these recurring anchors and stores the generated Gaussian data for each view. Materialization work thus grows with the total number of per-view anchor requests, even when the number of distinct requested anchors is much smaller. This repeated decoding and storage leaves the substantial overlap between neighboring views unexploited.
    \item \textbf{\textit{GPU-only Execution}:}
    Existing pipelines bind selection, materialization, and rendering to each view. Even when multiple views run concurrently, all three steps use GPU, while CPU cores remain unused for selection. Yet selecting anchors for a known pose depends only on that pose and the read-only scene, so selections for different views can be computed independently. Keeping these queries inside the per-view GPU pipeline limits the processing resources available to selection and delays the materialization that consumes its results.
    
\end{enumerate}

To eliminate these sources of waste, we propose a \uline{G}aussian \uline{D}ata \uline{M}anagement framework for \uline{G}aussian \uline{S}platting, referred to as \textbf{GDM-GS}. GDM-GS decouples anchor selection, Gaussian materialization, and image rendering, enabling optimizations within each view, across neighboring views, and across CPU and GPU resources. It reduces unnecessary anchor selection and repeated Gaussian materialization while preserving view-specific projection, depth ordering, and blending. However, these optimizations introduce three key challenges.

\textbf{Early Anchor Filtering}. Selection operates on stored anchors, but their relevance depends on the Gaussians they generate. These Gaussians can extend beyond the anchor location, and their exact shapes depend on decoder outputs. Therefore, only indexing anchors can miss visible content, while decoding and projecting every candidate would retain the cost we seek to eliminate. To enable early filtering without discarding potentially visible Gaussians, we propose the conservative spatial bounds from stored anchor geometry and decoder scale constraints. Then we build index on these bounds so that a spatial range query can reject entire groups of anchors without testing or projecting them individually. In addition, anchors inside the viewing region may be hidden behind nearer surfaces, which makes them useless and redundant for the rendering. However, their bounds cannot establish such occlusion because they may contain empty space. To exclude these hidden anchors before decoding, we build a separate index over solid regions derived from the scene mesh and remove the space they block from the spatial range query. The resulting query can prune the hidden anchors, skipping their descendant tests and preventing unnecessary Gaussian materialization (\S\ref{sec:Index}).

\textbf{Cross-View Materialization Sharing}. Neighboring views often select the same anchors, creating opportunities to avoid repeated decoding. However, the same anchor can generate different Gaussian attributes under different viewpoints, so directly reusing decoded data may degrade rendering quality. Moreover, neighboring views need not select identical anchors, and sharing one view’s selection may omit content needed by another. To preserve all requested anchors while eliminating repeated decoding, we group neighboring views, form the union of their selected anchors, and decode each distinct anchor once at a shared source pose. We use short groups and place the source near the group midpoint to limit viewpoint differences, with group length controlling the trade-off between reuse and fidelity. To preserve individual selections without duplicating the generated data, a membership map lets each view access its requested rows from a single shared copy. Each view then projects and renders those rows under its own pose, reducing repeated materialization and storage while retaining per-view rendering (\S\ref{sec:Cache}).

\textbf{CPU-GPU Parallel Execution.} Existing pipelines bind selection, materialization, and rendering to each view, so even concurrently processed views perform all three steps on the GPU. Our spatial-query formulation makes selection independent of Gaussian generation and other views, allowing CPU cores to participate. Exploiting this opportunity requires reorganizing execution while ensuring that all selections needed by a group are available before shared materialization. We therefore separate execution into two stages: the CPU and GPU first process a batch of selection queries concurrently; after all selections finish, the GPU materializes each group’s anchor union once and renders its views from the shared Gaussians. However, this stage boundary makes selection completion depend on the slower processor. To reduce waiting, we model effective CPU and GPU selection times (including result transfer) and allocate queries to minimize their maximum predicted completion time. This schedule combines parallel selection with grouped materialization (\S\ref{sec:Schedule}).

Our contributions are summarized as follows:
\begin{itemize}
    \item  We develop spatial indexes for anchor selection and occlusion-aware pruning, allowing irrelevant and fully occluded anchors to be filtered before Gaussian materialization.
    \item We introduce cross-view Gaussian sharing that materializes recurring anchors once for neighboring views while preserving each view's individual anchor requests.
    \item We design a CPU-GPU execution scheme that parallelizes anchor selection across processors and schedules queries using a cost model.
    \item We integrate these techniques into \textbf{GDM-GS}. Experiments on eight real-world datasets show that GDM-GS achieves up to $2\times$ faster trajectory rendering than state-of-the-art 3DGS methods, approaching the theoretical speedup bound.
\end{itemize}

%% file: sections/preliminary.tex
\section{Preliminaries}
\label{sec:prelim}
In this section, we first view 3DGS scenes as spatial data (\S\ref{sec:prelim:structured_3dgs}), then describe how a single view is rendered from them (\S\ref{sec:prelim:view}), and finally define trajectory rendering and our problem (\S\ref{subsubsec:prelim_stages_Rendering}).

\subsection{3DGS as Spatial Data}
\label{sec:prelim:structured_3dgs}
We first define the vanilla 3DGS, as illustrated in Figure \ref{fig:3DGS}-(a).

\begin{definition}[\textbf{3DGS Scene Representation}]
\label{def:3DGS}
3DGS scene is set of Gaussian primitives $\mathcal G=\{g_j\}_{j=0}^{N_g-1}$, where each primitive $g_j=(\mu_j,\Sigma_j,c_j,\alpha_j)$ is an ellipsoidal data unit has a center $\mu_j\in\mathbb{R}^3$, a covariance $\Sigma_j\in\mathbb{R}^{3\times3}$ that controls its size, shape, and rotation, a view-dependent color $c_j$, and an opacity $\alpha_j\in[0,1]$.
\end{definition}

From a spatial data management perspective, only $\mu_j$ and $\Sigma_j$ matter, as together they describe an ellipsoid that bounds the Gaussians in space; $c_j$ and $\alpha_j$ are payload that only the final blending reads. Therefore, rendering a view is essentially a spatial range query: it retrieves the ellipsoids that intersect the region visible from the camera and then combines their payload. However, vanilla 3DGS stores all $N_g$ Gaussians with complete attributes, which leads to prohibitive storage and I/O on large scenes with millions to billions of primitives. To mitigate this, Structured 3DGS~\cite{lin2024scaffoldgs,ren2024octree} only stores a sparse set of \textit{anchors} and derives the Gaussians from them on demand, as illustrated in Figure~\ref{fig:3DGS}-(b) and defined below:

\begin{figure}
    \centering
    \includegraphics[width=1\columnwidth]{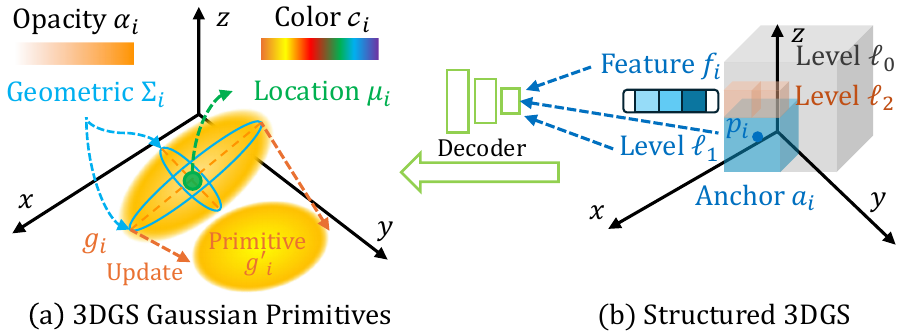}
    \caption{3DGS Data Representation}
    \label{fig:3DGS}
\end{figure}

\begin{definition}[\textbf{Structured 3DGS Scene Representation}]
\label{def:Structured_3DGS}
A structured 3DGS scene $(\mathcal{A},\theta)$ consists of an anchor set $\mathcal{A}=\{a_i\}_{i=0}^{N_a-1}$ and the parameters $\theta$ of a shared decoder $\mathrm{Dec}_\theta$. Each anchor $a_i=(p_i,s_i,\ell_i,f_i)$ has a position $p_i\in\mathbb{R}^3$, a grid size $s_i$ initialized to the voxel size of its level of detail (LoD) $\ell_i$, and a latent feature $f_i$. It also stores $n_o$ learned offsets, fixed after reconstruction, which $s_i$ scales into displacements $\Delta_{i,k}$; its $k$-th candidate Gaussian is centered at $\mu_{i,k}=p_i+\Delta_{i,k}, \forall k\in[0,n_o)$. Given a camera center $\pi$, $\mathrm{Dec}_\theta$ maps $f_i$ and the viewing direction and distance from $\pi$ to $p_i$ to the covariance, color, and opacity of each candidate, with a size bounded by $s_i$.
\end{definition}

Again, only a small part of an anchor is spatial. Since the offsets are fixed and expressed in units of $s_i$, and every generated Gaussian is sized by at most $s_i$, the position $p_i$ and the grid size $s_i$ alone determine where the Gaussians of $a_i$ can lie, independently of the pose and the decoder. The feature $f_i$ and the decoder $\theta$ are payload needed only to decode, and $\ell_i$ is a level used for distance-based filtering. Therefore, the anchor table is a compact spatial table of $N_a$ objects managed through $(p_i,s_i)$, while each anchor stands for at most $n_o$ Gaussians; we call $N_a n_o$ the \emph{learned Gaussian capacity} of the scene. Two consequences shape the rest of the paper: retrieval can be answered at the anchor level before any decoding, and the Gaussian level must be produced by an expensive decode before anything can be rendered.

\subsection{3DGS Rendering Pipeline}
\label{sec:prelim:view}
After describing what a scene stores, we now describe how to render a view (image) from it. A scene is first reconstructed offline from training images with known camera poses; reconstruction optimizes $\mathcal{A}$ and $\theta$ and may insert or delete anchors along the way. Rendering then treats the result as a fixed, read-only database, in which every anchor keeps a stable ID.

Each view is requested by a \textit{\textbf{camera pose}} $\pi=(R,t)$, which maps a world point $x$ to camera coordinates $Rx+t$ with a rotation $R\in\mathbb{R}^{3\times3}$ and a translation $t\in\mathbb{R}^3$; its camera center is $o(\pi)=-R^\top t$. Rendering the view is essentially a spatial range query, and its query region is the \textbf{\emph{view frustum}} $\mathcal{F}(\pi)$: the points that project into the image and lie beyond the \emph{near plane}, \ie the minimum depth that the renderer draws. Existing structured 3DGS methods~\cite{lin2024scaffoldgs,ren2024octree} answer this query in three steps, as shown in Figure~\ref{fig:3DGS}:
\begin{enumerate}
    \item[S1)] \textbf{\emph{Anchor Selection}}: $S(\pi)=\mathrm{Select}(\mathcal{A},\pi)$ returns the IDs of the anchors that pass the distance-based LoD rule and an anchor-level test against $\mathcal{F}(\pi)$. In the baseline, this is a full scan of the anchor table with a per-row geometric predicate.
    \item[S2)] \textbf{\emph{Gaussian Materialization}}: $\mathcal{G}(\pi)=\mathrm{Dec}_\theta(S(\pi),\pi)$ decodes the selected anchors under $o(\pi)$ and keeps the candidates with positive opacity, so $|\mathcal{G}(\pi)|\le|S(\pi)|\,n_o$.
    \item[S3)] \textbf{\emph{Image Rendering}}: $\hat{I}(\pi)=\mathrm{Render}(\mathcal{G}(\pi),\pi)$ applies Gaussian-level tests, projects the surviving Gaussians onto the image plane, sorts them by depth, and composites the image.
\end{enumerate}
Two features of this pipeline matter for efficiency. First, filtering happens twice: coarsely per anchor in S1 before decoding, and finely per Gaussian in S3 after decoding, so a Gaussian rejected in S3 has already paid for its decoding in S2. Second, all three steps run on the GPU, one after another, so the CPU stays idle and no step overlaps with another.

\begin{figure}[t]
    \centering
    \includegraphics[width=1\columnwidth]{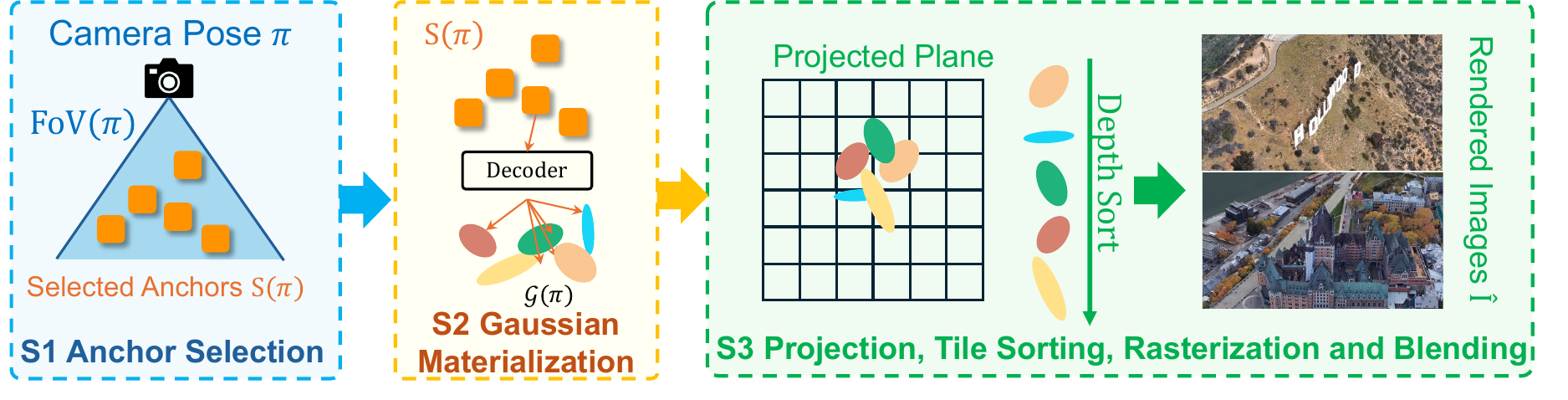}
    \caption{\textbf{Vanilla Structured 3DGS Rendering Pipeline.}}
    \label{fig:vanillapipe}
\end{figure}

\subsection{Trajectory Rendering Problem}
\label{subsubsec:prelim_stages_Rendering}
Section~\ref{sec:prelim:view} renders a single view. A trajectory instead requests many views in a given order:

\begin{definition}[\textbf{Trajectory Rendering}]
\label{def:Trajectory_Rendering}
Given a scene $(\mathcal{A},\theta)$ and a trajectory $\Pi=(\pi_1,\ldots,\pi_T)$ of $T$ target poses, trajectory rendering outputs the images $(\hat{I}_1,\ldots,\hat{I}_T)$ in trajectory order. For target $t$, $S_t=S(\pi_t)$ is its selected set with $K_t=|S_t|$, $\mathcal{G}_t$ is its Gaussian input to S3, and $\hat{I}_t=\mathrm{Render}(\mathcal{G}_t,\pi_t)$. Execution across targets may overlap.
\end{definition}

We write $\mathcal{G}_t$ separately from $\mathcal{G}(\pi_t)$ because the two differ once decoding is shared across targets (Section~\ref{sec:Cache}). Unlike isolated views, the targets of a trajectory are not independent queries, and they have two properties:
\begin{enumerate}
    \item \textbf{Query Locality.} Camera trajectories are often smooth at frame granularity, so consecutive frusta overlap, and consecutive targets request largely the same anchors. Since decoding is view-dependent, the same anchor still yields slightly different attributes under different poses, so this overlap permits approximate rather than exact reuse.
    \item \textbf{Advance Knowledge.} The whole trajectory is given before rendering starts, and $S_t$ depends only on $\pi_t$, not on earlier images. Hence, work for later targets can start before earlier targets finish.
\end{enumerate}

Existing structured 3DGS methods nevertheless render a trajectory view by view: they run the pipeline of Section~\ref{sec:prelim:view} for each target independently and in order, and thus exploit neither property. We call this straightforward strategy the \emph{reference execution} and use it as the baseline throughout the paper. Its outputs, marked with a star as $(S^\star_t,\mathcal{G}^\star_t,\hat{I}^\star_t)$, are the images that a faster strategy should reproduce, and its latency $T^\star_\Pi=\sum_{t=1}^{T}\big(T^\star_{\mathrm{sel},t}+T^\star_{\mathrm{mat},t}+T^\star_{\mathrm{rend},t}\big)$ is the time it should beat, where the three terms are the latencies of S1, S2, and S3 for target $t$. Since the reuse enabled by Query Locality is approximate, we require bounded distortion rather than identical images, which leads to our problem:

\begin{definition}[\textbf{3DGS Trajectory Rendering Problem}]
\label{def:Problem_Definition}
Given a structured 3DGS scene $(\mathcal{A},\theta)$ and a trajectory $\Pi$, output $(\hat{I}_1,\ldots,\hat{I}_T)$ in trajectory order with the minimum wall-clock latency $T_\Pi$, while keeping the quality of the reference execution, i.e., $\mathcal{L}_{\mathrm{img}}(\hat{I}_t,\hat{I}^\star_t)\le\varepsilon$ for every target $t$, where $\mathcal{L}_{\mathrm{img}}$ is an image distortion measure and $\varepsilon$ is a small threshold.
\end{definition}

Here, $T_\Pi$ runs from the start of the trajectory until its last image is collected. Unlike $T^\star_\Pi$, it need not be a sum of per-target latencies, since the steps of different targets may overlap.

%% file: sections/Overview.tex
\section{GDM-GS Framework}
\label{sec:Overview}
Definition~\ref{def:Problem_Definition} asks for a latency below $T^\star_\Pi$ at nearly the reference quality. To this end, we first identify where the reference execution wastes latency, and then present GDM-GS, a data-management framework that removes each waste by decoupling the S1--S3 steps that the reference execution runs together.

\textbf{Optimization Opportunities.} Combining the two pipeline features of Section~\ref{sec:prelim:view} with the two trajectory properties of Section~\ref{subsubsec:prelim_stages_Rendering}, the reference execution wastes latency at three levels, each tied to terms of $T^\star_\Pi$:
\begin{enumerate}
    \item[O1)] \emph{\textbf{Within a Target:}} S1 is the most time-consuming step. It scans all $N_a$ anchors, and it still passes to S2 anchors whose Gaussians are later rejected in S3 or hidden behind nearer surfaces, each costing $n_o$ decodes. A more selective S1 could reduce both the selection and the materialization terms.
    \item[O2)] \emph{\textbf{Across Targets:}} By \textbf{\textit{Query Locality}}, consecutive targets decode largely the same anchors. Decoding each requested anchor once for several targets reduces the materialization term, at the price of approximating per-target decoding.
    \item[O3)] \emph{\textbf{Across Processors:}} By \textbf{\textit{Advance Knowledge}}, all poses are known in advance and each $S_t$ depends only on $\pi_t$, so the selections of many targets can be answered before any of them is decoded. Yet the reference execution answers every S1 on the GPU, one target after another, while the CPU stays idle. Once S1 no longer needs the GPU, answering these selections on both processors shortens the selection term.

\end{enumerate}
S3 stays per target in all three cases, since projection and blending depend on each pose.

\textbf{Framework Overview.} GDM-GS exploits the three opportunities with three components, one per level, each answering a classical data management question: how to index, how to share, and how to schedule. First, \S\ref{sec:Index} exploits O1 by indexing: an anchor index answers S1 as a spatial range query instead of a full scan, and an occluder index cuts holes into the query region for anchors hidden behind surfaces, so fewer anchors reach S2. Since this query needs neither projection nor decoding, it also frees S1 from the GPU, which O3 requires. Second, \S\ref{sec:Cache} exploits O2 by sharing: it merges the requests of a group of consecutive targets and decodes each distinct anchor once under a shared source pose, while each target still performs S3 under its own pose. Third, \S\ref{sec:Schedule} exploits O3 by scheduling: it answers the selections of a batch of groups on both the CPU and the GPU, split by their measured throughput, and then materializes and renders the batch's groups on the GPU under a bounded budget of shared Gaussian rows. For each batch, GDM-GS thus runs S1 once per target on either processor, S2 once per group, and S3 once per target. Together, the three components reduce unnecessary inputs, eliminate repeated materialization, and shorten selection, while preserving the S1--S3 dependencies of each target.

%% file: sections/method.tex
\begin{figure*}[t]
    \centering
    \includegraphics[width=1\linewidth]{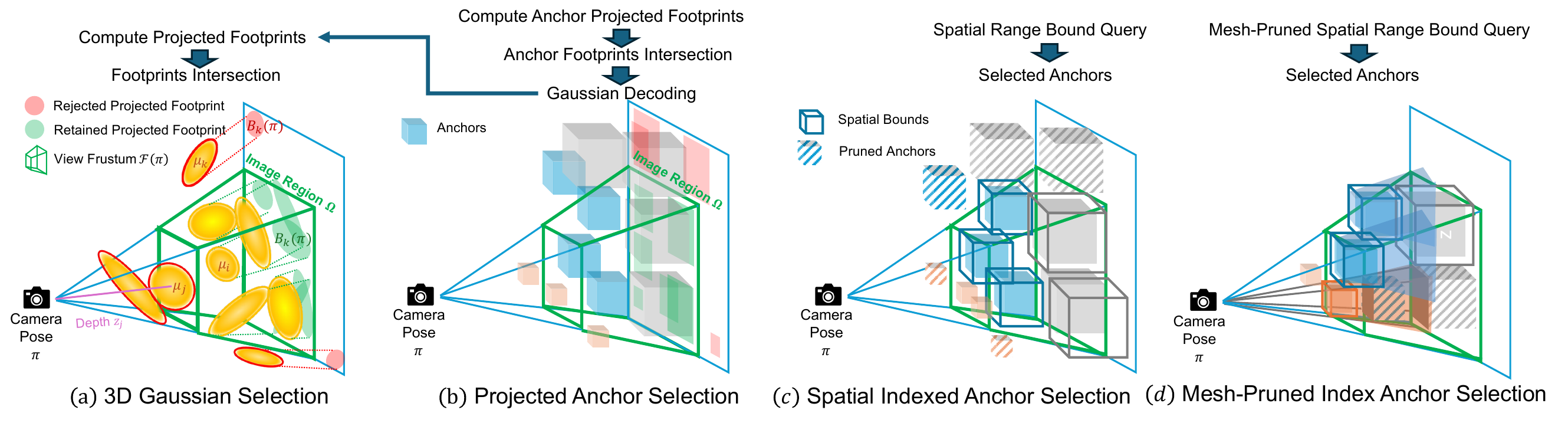}
    \caption{View-dependent selection for 3D Gaussian splatting}
    \label{fig:queryalgo}
\end{figure*}

\section{Efficient Gaussian Selection}
\label{sec:Index}
Section~\ref{sec:prelim:structured_3dgs} showed that rendering a view is a spatial range query over the scene. However, existing 3DGS pipelines never answer it as one: vanilla 3DGS projects every Gaussian and structured 3DGS projects every anchor, both by brute force on the GPU and as part of rendering. This section introduces spatial data management into this step to reduce the cost. We first review the two brute-force procedures in detail (\S\ref{subsec:Index_Basic} and \S\ref{subsubsec:Index_Anchor_Existing}), then replace them with a spatial index over anchor bounds (\S\ref{subsubsec:Index_Anchor_Index}), and finally cut occluded space out of the query with an occluder index (\S\ref{subsec:Index_Mesh}).

\subsection{Standard 3DGS Gaussian Selection}
\label{subsec:Index_Basic}
As illustrated in Figure~\ref{fig:queryalgo}-(a), given the Gaussian table $\mathcal{G}=\{g_j\}$ and a pose $\pi$, 3DGS selects the primitives that may contribute to the image, \ie those whose ellipsoids meet the frustum $\mathcal{F}(\pi)$. Because an ellipsoid centered outside the frustum may still reach into it, testing the center $\mu_j$ alone is insufficient. The renderer therefore evaluates the overlap on the image plane with two quantities: 1) the depth $z_j(\pi)$ of $\mu_j$ along the viewing direction, and 2) the projected footprint $B_j(\pi)$, a bounded 2D region that represents the extent of the ellipsoid on the image plane. With $\Omega$ the image domain and $z_{\min}$ the depth of the near plane, the selection predicate is $P(g_j,\pi)=\big[z_j(\pi)>z_{\min}\big]\wedge\big[B_j(\pi)\cap\Omega\neq\emptyset\big]$.

The procedure evaluates $P$ record by record. It first tests the inexpensive depth condition and discards the records that fail it; for each remaining record, it computes the footprint and tests its intersection with $\Omega$; finally, it compacts the qualifying IDs into a candidate sequence. In Figure~\ref{fig:queryalgo}-(a), $\mu_i$ and the other green footprints qualify, $\mu_j$ fails the depth test and is discarded without projection, and $\mu_k$ and the other red footprints are projected but miss $\Omega$.
 
\textbf{Complexity Analysis.} Each test takes constant time, so selection takes $\Theta(N_g)$ work and $O(N_g)$ space for the mask and the output. From a data-management perspective, this is a full-table scan whose per-row predicate is a pose-dependent projection. No access method is involved: every record undergoes the depth test, and every depth-qualified record pays for its footprint even when it is rejected (red footprints in (a)). It should be noted that this scan is not a separate query: it is fused into the preprocessing of the GPU rasterizer, so spatial filtering is a by-product of rendering.

\subsection{Efficient Spatial-Based Anchor Selection}
\label{subsec:Index_Anchor}
The Gaussian-level scan examines $\Theta(N_g)$ primitives individually, although most of them lie outside the view. Structured 3DGS offers a coarser selection unit, the anchor. We first review how existing methods use it, and then show how we index it.

\subsubsection{\textbf{Existing Projected Anchor Selection}}
\label{subsubsec:Index_Anchor_Existing}
As introduced in \S\ref{sec:prelim:structured_3dgs}, an anchor stands for up to $n_o$ Gaussians, so a group of candidate Gaussians can be excluded together before decoding. As shown in Figure~\ref{fig:queryalgo}-(b), it first uses $p_i$ and $\ell_i$ for the LoD rule, admitting finer levels in nearby regions and coarser levels in distant ones. This $O(N_a)$ scan decides LoD eligibility, not geometric visibility. Eligible anchors then undergo two levels of geometric filtering, separated by Gaussian materialization:

\noindent\underline{\textit{Step 1: Anchor-proxy filtering before decoding}}. In S1, the stored geometry of each eligible anchor defines a lightweight proxy, whose projected extent is tested against $\Omega$. This costs $O(1)$ per anchor and $O(N_a)$ in total. The survivors form $S_t$ with $K_t=|S_t|$ and enter S2, which decodes $\Theta(K_t n_o)$ anchors into $|\mathcal{G}_t|\le K_t n_o$ Gaussians.

\noindent\underline{\textit{Step 2: Gaussian-level filtering after decoding}}. In S3, each decoded Gaussian is tested with the predicate $P$ of Section~\ref{subsec:Index_Basic}, which costs $O(|\mathcal{G}_t|)$. Passing the anchor proxy does not guarantee passing $P$, so a Gaussian rejected here has already paid for its decoding.

Together, the two levels cost $O(N_a+|\mathcal{G}_t|)$ geometric work, which is less than $\Theta(N_g)$ because the scan runs over fewer records. Nevertheless, the method remains a brute-force scan on the GPU: every anchor, including those far outside the view, is projected at every frame, and filtering still happens inside the rendering pipeline rather than as a query over stored data. Thus, neither vanilla nor structured 3DGS manages its primitives as spatial data.

\begin{figure*}[t]
    \centering
    \includegraphics[width=1\linewidth]{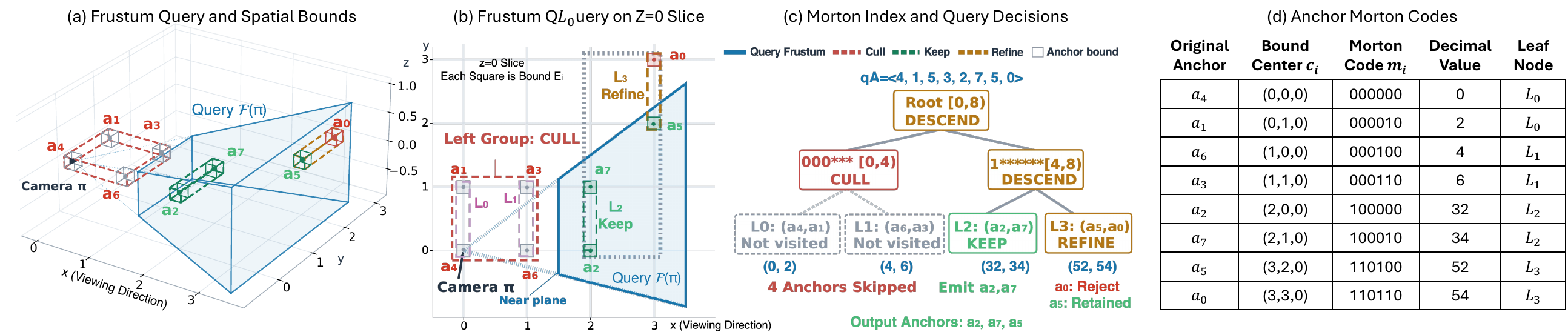}
    \caption{Anchor Index and Frustum Query Example}
    \label{fig:index}
\end{figure*}

\subsubsection{\textbf{Spatial Index-Based Anchor Selection}}
\label{subsubsec:Index_Anchor_Index}
We take the opposite view: S1 is a range query over the anchor table, so it should be answered by a spatial index rather than by projecting every record. To the best of our knowledge, no existing 3DGS pipeline answers selection this way. Two obstacles stand in the way: 1) The per-row predicate is a pose-dependent projection, which no index can store; 2) Even an indexable predicate would still be evaluated on all $N_a$ records by a scan. We remove the two obstacles in turn: Reduction~1 replaces the projection with a pose-independent key, and Reduction~2 indexes these keys.

\textbf{\underline{\textit{Reduction 1}}: \textit{Selection without Decoding or Projection}.} An anchor $a_i$ is not itself a Gaussian: its Gaussians are displaced from $p_i$ and have spatial extent, so testing $p_i$ alone may miss a displaced Gaussian that reaches the frustum. However, Section~\ref{sec:prelim:structured_3dgs} shows that $p_i$ and $s_i$ alone determine where these Gaussians can lie. Specifically, the $n_o$ candidate centers $\mu_{i,k}=p_i+\Delta_{i,k}$ are known without decoding, and we enclose them in the axis-aligned box $C_i=[l_i,u_i]$, whose corners are their coordinate-wise minima and maxima. Let $\bar r_i$ be a conservative bound on the world-space radius of any Gaussian of $a_i$, given by $s_i$ and the renderer's finite support. The indexed bound of $a_i$ is $E_i=\big[\,l_i-\bar r_i\mathbf{1},\ u_i+\bar r_i\mathbf{1}\,\big], \mathbf{1}=(1,1,1)^\top$.

Unlike a projected footprint, $E_i$ is pose-independent and computed once per scene. An anchor whose bound misses $\mathcal{F}(\pi)$ can be discarded in S1 without decoding or projecting anything; surviving anchors still proceed to S2 and to the exact tests of S3, so $E_i$ is a filter rather than a replacement for them.

\textit{\underline{\textbf{Reduction 2}}\textbf{: Indexed Spatial Range Query}}. 
A scan over $\{E_i\}$ still examines all $N_a$ records. We therefore treat the bounds as the keys of a 3D range query, $Q(\pi)=\{\, i \mid E_i\cap\mathcal{F}(\pi)\neq\emptyset \,\}$, answered by an index that is built once per scene, while each pose supplies a new query region.

\textbf{Index Structure.} To pack nearby bounds into contiguous storage, we sort the anchors by the Morton code $m_i$ of their bound centers $(l_i+u_i)/2$, obtained by normalizing the coordinates to the scene extent, quantizing them, and interleaving their bits. The sorted IDs form the packed array $q_A$, which we split into leaves of at most $b$ consecutive IDs. A binary radix tree groups the leaves by the longest shared prefix of their first keys, like a linearized R-tree laid out in Morton order. Each node $v$ stores its two children, the interval $[s_v,e_v)$ of $q_A$ covered by its subtree, and a group bound $E_v$ that contains every descendant bound. Morton order thus decides which anchors are grouped and makes every subtree one contiguous ID range, whereas all geometric tests use the real-valued bounds. Figure~\ref{fig:index}-(a,c,d) shows eight anchors on a two-bit grid, sorted into $q_A=\langle 4,1,6,3,2,7,5,0\rangle$ and packed into leaves $L_0,\ldots,L_3$ with $b=2$.

Computing the bounds takes $\Theta(N_a n_o)$ time, encoding $\Theta(N_a)$, sorting $O(N_a\log N_a)$, and building the tree with its group bounds $\Theta(N_a)$, so construction takes $O(N_a(n_o+\log N_a))$ once per scene. The index stores $q_A$, the bounds, and $O(N_a/b)$ nodes of constant size, \ie $\Theta(N_a)$ space, independent of $n_o$ and of the pose.

\textbf{Query Processing.} The query walks the tree from the root and classifies each group bound $E_v$ by box--plane tests against the constantly many planes of $\mathcal{F}(\pi)$. For a box with center $c$ and half-widths $h$ and an inward plane $n^\top x+d\ge 0$, every point of the box has a signed value within $[\,n^\top c+d-|n|^\top h,\ n^\top c+d+|n|^\top h\,]$, so each test takes $O(1)$ time. The three outcomes are the standard spatial-index decisions of Algorithm~\ref{alg:frustum-query}:
\begin{itemize}
    \item \textbf{Cull:} $E_v$ lies outside some plane, so the subtree is discarded.
    \item \textbf{Keep:} $E_v$ lies inside every plane, i.e., $E_v\subseteq\mathcal{F}(\pi)$, so the interval $q_A[s_v:e_v)$ is reported without testing any descendant.
    \item \textbf{Descend:} otherwise, the walk recurses; a leaf tests each $E_i$.
\end{itemize}
In Figure~\ref{fig:index}-(b,c), one Cull skips the four anchors of the left subtree, Keep reports $L_2=(2,7)$ as an interval, and only $L_3$ is refined, returning $\langle 2,7,5\rangle$. Let $N^{\mathrm{node}}_t$ and $N^{\mathrm{refine}}_t$ be the numbers of classified nodes and refined anchors for $\pi_t$. The query takes $O(N^{\mathrm{node}}_t+N^{\mathrm{refine}}_t+|Q(\pi_t)|)$ time. It degrades to $O(N_a)$ only if the frustum meets every bound, whereas in trajectories the field of view is small compared with the scene, so Cull and Keep typically resolve most of the tree near the root. Unlike the brute-force scans of \S\ref{subsec:Index_Basic} and \S\ref{subsubsec:Index_Anchor_Existing}, the cost thus follows how much of the scene the view covers rather than the size of the scene.

\textbf{Correctness.} As for any database range index, we separate two statements: the keys are a conservative surrogate of the true predicate, and the tree answers the key predicate exactly.
 
\begin{lemma}[\textbf{Filter Completeness}]
\label{lem:filter}
If any Gaussian that $a_i$ can generate intersects $\mathcal{F}(\pi)$, then $E_i\cap\mathcal{F}(\pi)\neq\emptyset$, hence $i\in Q(\pi)$.
\end{lemma}
\begin{proof}
Every generated center lies in $C_i$, and every generated support lies in its $\bar r_i$-expansion $E_i$, so an intersecting support meets $E_i\cap\mathcal{F}(\pi)$.
\end{proof}
 
\begin{lemma}[\textbf{Exact Range Query}]
\label{lem:exact}
Alg.~\ref{alg:frustum-query} returns exactly $Q(\pi)$.
\end{lemma}
\begin{proof}
Cull is sound because $E_v$ contains every descendant bound; Keep is complete because every descendant satisfies $E_i\subseteq E_v\subseteq\mathcal{F}(\pi)$; and Descend partitions the ID range, with leaves testing each record.
\end{proof}
 
Together, the two lemmas give the filter-and-refine guarantee that S1 needs: no contributing anchor is lost, and anchors whose bounds miss the frustum are never decoded. The remaining false positives come from conservative bounds, which S3 resolves, and from anchors hidden behind surfaces, which we address next. %

\begin{algorithm}[t]
\caption{Frustum Index Query}
\label{alg:frustum-query}
\small
\KwIn{Index $I_A=(q_A,\textit{tree})$; pose $\pi$ with planes of $\mathcal{F}(\pi)$}
\KwOut{ ID sequence of $Q(\pi)=\{i\mid E_i\cap\mathcal{F}(\pi)\neq\emptyset\}$}
\SetKwProg{Function}{Function}{:}{}
\SetKwFunction{FrustumWalk}{FrustumWalk}

Initialize an empty output list $O$\;
\FrustumWalk{\textit{root}, $\pi$, $O$}\;
\Return $O$\;
\Function{\FrustumWalk{$v,\pi,O$}}
{
    Test $E_v$ against the planes of $\mathcal{F}(\pi)$ \Comment{$O(1)$ Box--Plane tests}\;
    \If{Cull}
    {
        \Return \Comment{Prune the whole subtree}\;
    }
    \ElseIf{Keep}
    {
        Append $q_A[s_v:e_v)$ to $O$; \Return\;
    }
    \ElseIf{$v$ is a leaf}
    {
        \For{each original ID $i$ in $q_A[s_v:e_v)$}
        {
            \If{$E_i\cap\mathcal{F}(\pi)\neq\emptyset$}
            {
                Append $i$ to $O$;
            }
        }
        \Return\;
    }
    \FrustumWalk{left child of $v$, $\pi$, $O$}\;
    \FrustumWalk{right child of $v$, $\pi$, $O$}\;
}
\end{algorithm}

\subsection{Occluder-Pruned Anchor Selection}
\label{subsec:Index_Mesh}
$Q(\pi)$ still contains anchors that lie in the frustum but behind nearer surfaces, as illustrated in Figure~\ref{fig:queryalgo}-(d), and their Gaussians contribute little or nothing to the image. Vanilla and structured 3DGS do not filter such Gaussians at all: they are decoded, projected, and sorted, and only alpha blending makes them invisible. Proxy-GS~\cite{proxygs2026} uses mesh-based occlusion to skip decoding such anchors, but they are still selected and tested one by one. We instead remove occluded space from the query region itself: \S\ref{subsubsec:Index_Mesh_Cell} builds an occluder index over the solid space of the scene, and \S\ref{subsubsec:Index_Mesh_Query} uses it to cut holes into the frustum before the anchor index is searched.

\subsubsection{\textbf{Occluder Index}}
\label{subsubsec:Index_Mesh_Cell}
The anchor bounds cannot serve as occluders, because $E_i$ may contain empty space, and whether its Gaussians block anything depends on decoded attributes, which we aim not to compute. The reconstructed mesh, stored as a vertex table $V$ and a triangle table $F$ with $N_m=|F|$, does describe the blocking surfaces. However, a single triangle can hardly hide a whole $E_i$, and testing $N_m$ triangles against every bound would cost more than the decoding it saves. Therefore, we manage the space the triangles enclose. The surface separates \emph{free space}, from which the cameras observe the scene, and \emph{solid space} inside objects, which we treat as opaque. Since the mesh and the anchors come from the same point cloud, we discretize solid space on the Morton grid of \S\ref{subsubsec:Index_Anchor_Index}, using its cells at a fixed level $L$ as a leaf grid of $G=8^L$ cells.

\begin{definition}[\textbf{Occluder Cell}]
\label{def:occ-cell}
A cell is an \emph{occluder cell} if it lies entirely in solid space.
\end{definition}
Leaf cells are classified in two passes. Every cell met by a triangle is a \emph{boundary cell} and never an occluder cell; a flood fill from the cells of the training camera centers, which lie in free space, then marks the reachable non-boundary cells as free and the unreachable ones as occluder cells. A gap in the mesh can only let the flood fill leak into solid space and turn occluder cells into free ones; it never marks a cell reachable from the cameras as solid. This mirrors the anchor bounds:
\begin{property}[\textbf{Two-Sided Conservativeness}]
\label{prop:two_sided}
An anchor bound over-approximates the space its Gaussians may occupy and is used only to decide what is visible; an occluder cell under-approximates the space that blocks the view and is used only to decide what is hidden. An error on either side keeps more anchors, never fewer.
\end{property}

However, a leaf cell is small, and a bound typically straddles several cells, each hiding only part of it. We therefore merge occluder cells bottom-up in an octree on the same grid, where the eight children of a node $u$ split its cell $c_u$ in half along each axis.
 
\begin{lemma}[\textbf{Hierarchical Occluder Merge}]
\label{lem:occ_merge}
A node $u$ is an occluder cell if and only if all eight children of $u$ are occluder cells.
\end{lemma}
\begin{proof}
The eight children partition $c_u$ exactly, so $c_u$ lies in solid space if and only if every child does.
\end{proof}

Merging thus produces cells large enough to hide a whole bound, while a mixed node is never an occluder cell, so no incomplete solid is used. The result is a region octree, denoted $I_O$, whose nodes are \emph{full} (an occluder cell), \emph{empty} (no occluder cell inside), or \emph{partial}. Full nodes are stored as leaves, empty nodes are omitted, and partial nodes store their non-empty children. Nodes are addressed by Morton prefixes, so $I_O$ shares the coordinate system of the anchor index while remaining a separate index. Construction marks boundary cells per triangle and visits each leaf cell once in the flood fill and the merge, taking $O(N_m+G)$ time when each triangle meets a bounded number of cells. Since every partial node contains a boundary cell and every stored full node has a partial parent, $I_O$ stores $O(N_b L)$ nodes for $N_b$ boundary cells, which grows with the mesh area rather than with the enclosed volume.

\subsubsection{\textbf{Holed-Frustum Query}}
\label{subsubsec:Index_Mesh_Query}
With $I_O$ in place, a pose is answered in two steps: 1) we first retrieve the occluder cells in the frustum and turn each of them into a hole in the query region, and then 2) query the anchor index with this holed frustum. Since all holes are known before the anchor index is searched, every decision of the second step is final.

\textbf{Step 1: Occluder Retrieval.} This step is a frustum range query on $I_O$ with the box--plane tests of Section~\ref{subsubsec:Index_Anchor_Index}: a node whose cell misses $\mathcal{F}(\pi)$ is culled, a partial node is descended, and a full node is reported as the coarsest occluder cell on its path. Since the renderer clips everything in front of the near plane, a full cell that crosses this plane, e.g., one containing the camera center, may not block the view; it is split into its eight children, which are full by Lemma~\ref{lem:occ_merge}, and leaf cells that still cross the plane are skipped. The output is the set $C(\pi)$ of occluder cells that meet the frustum and lie beyond the near plane.

\begin{definition}[\textbf{Hole}]
\label{def:hole}
For a cell $c\in C(\pi)$, the \emph{hole} $\mathcal{H}_c(\pi)$ is the set of points $p$ such that the segment from $o(\pi)$ to $p$ passes through $c$ before reaching $p$. The \emph{holed frustum} is
$\mathcal{F}_{\mathrm{M}}(\pi)=\mathcal{F}(\pi)\setminus\bigcup_{c\in C(\pi)}\mathcal{H}_c(\pi)$.
\end{definition}

Since $o(\pi)$ lies outside $c$, the lines of sight through $c$ form a convex cone bounded by the planes through $o(\pi)$ and the silhouette edges of $c$, and a point of this cone lies behind $c$ exactly when it is on the inner side of every face of $c$ that faces the camera. Hence $\mathcal{H}_c(\pi)$ is a convex polyhedron with at most nine planes, like the frustum, and the same box--plane tests decide whether a box lies inside or outside it.
 
\begin{lemma}[\textbf{Hole Soundness}]
\label{lem:hole_sound}
If $E_i\subseteq\mathcal{H}_c(\pi)$ for some $c\in C(\pi)$, then no Gaussian of $a_i$ can contribute to the view.
\end{lemma}
\begin{proof}
By the construction of $E_i$ (Lemma~\ref{lem:filter}), every Gaussian of $a_i$ lies in $\mathcal{H}_c(\pi)$, so every line of sight to it passes through $c$, which is solid and beyond the near plane.
\end{proof}
 
Since a bound that only partially meets holes may still be visible, the anchors to retrieve are
\begin{equation}
\label{eq:qm}
Q_{\mathrm{M}}(\pi)=\big\{\, i \;\big|\; E_i\cap\mathcal{F}(\pi)\neq\emptyset\ \wedge\ \forall c\in C(\pi):\ E_i\not\subseteq\mathcal{H}_c(\pi) \,\big\}.
\end{equation}
We test holes one at a time rather than their union, whose coverage test is expensive. A bound hidden only jointly by several holes is therefore safely kept, and merging reduces such cases by making each hole as large as possible. Thus, $Q_{\mathrm{M}}(\pi)$ contains every anchor whose bound meets $\mathcal{F}_{\mathrm{M}}(\pi)$ and is contained in $Q(\pi)$.
 
\textbf{Step 2: Anchor Query on the Holed Frustum.} We answer Eq.~\eqref{eq:qm} with the walk of Algorithm~\ref{alg:frustum-query}, changing two decisions: a node is also culled if $E_v$ lies inside a single hole, since its whole subtree is then hidden by Lemma~\ref{lem:hole_sound}, and it is kept only if $E_v\subseteq\mathcal{F}(\pi)$ and $E_v$ meets no hole. Both decisions stop the search early. To avoid testing every hole at every node, the walk passes to the children of $v$ only the holes that $E_v$ is not outside of, since a hole that misses $E_v$ also misses every descendant bound. Keep then reduces to $E_v\subseteq\mathcal{F}(\pi)$ with an empty hole list, and deeper nodes test only nearby holes.
 
\begin{lemma}[\textbf{Safe Pruning}]
\label{lem:holed_exact}
Every anchor that the holed walk drops either lies outside $Q(\pi)$ or has its bound inside a retrieved hole. Hence, if solid space is opaque, the walk never drops an anchor that can contribute to the view.
\end{lemma}
\begin{proof}
A hole leaves the list at $v$ only when $E_v$ or an ancestor bound lies outside it, so removing it from the list drops no descendant. Cull by the frustum drops only IDs outside $Q(\pi)$, as in Lemma~\ref{lem:exact}. Cull by a hole $c$ drops only IDs with $E_i\subseteq E_v\subseteq\mathcal{H}_c(\pi)$, and a leaf drops an ID only if $E_i$ misses $\mathcal{F}(\pi)$ or lies inside a hole of its list. Every dropped ID of $Q(\pi)$ therefore lies inside a hole, and Lemma~\ref{lem:hole_sound} applies.
\end{proof}
 
It should be noted that Lemma~\ref{lem:holed_exact} guarantees safety, not completeness: the walk removes the hidden anchors that the retrieved holes can prove hidden, but some hidden anchors may survive. This happens when a bound is hidden only jointly by several holes, when the mesh misses or leaves open an occluding surface so that the flood fill marks the space behind it as free (Property~\ref{prop:two_sided}), or when an occluder is thinner than a leaf cell and thus yields only boundary cells. Such anchors are decoded and resolved by S3, exactly as in the baseline. Mesh quality therefore affects how many anchors are pruned, but not which visible anchors are kept.
 
\textbf{Complexity.} Step 1 visits $N^{\mathrm{occ}}_t$ nodes of $I_O$ with $O(1)$ work each. Step 2 visits $N^{\mathrm{node}}_t$ nodes and $N^{\mathrm{refine}}_t$ records of the holed walk, each of which additionally tests the holes in its list, so it takes $O\big((N^{\mathrm{node}}_t+N^{\mathrm{refine}}_t)(1+|C(\pi_t)|)\big)$ time in the worst case. Holes add a reason to Cull, which stops the search above hidden subtrees, but they narrow Keep, which may force a descent where a hole boundary crosses a bound; the output, however, only shrinks, since $Q_{\mathrm{M}}(\pi_t)\subseteq Q(\pi_t)$. Finally, S1 applies the LoD rule of \S~\ref{sec:prelim:view} to the retrieved IDs and returns $S(\pi_t)\subseteq Q_{\mathrm{M}}(\pi_t)$ to S2, as shown in Figure~\ref{fig:queryalgo}-(d).

\section{Shared Materialization}
\label{sec:Cache}
\begin{figure}[t]
    \centering
    \includegraphics[width=1\linewidth]{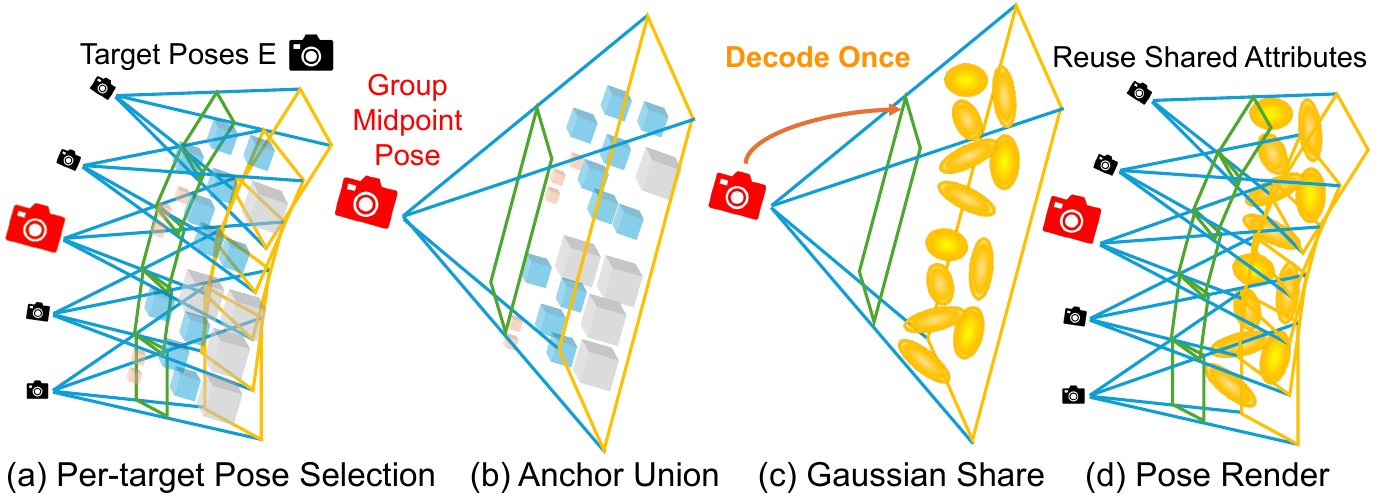}
    \caption{Overview of Shared Materialization.}
    \label{fig:cache}
\end{figure}

\S\ref{sec:Index} improves efficiency within a single view. However, consecutive poses of a trajectory observe largely the same scene content (Query Locality in \S\ref{subsubsec:prelim_stages_Rendering}), so their selected sets overlap, and the vanilla pipeline decodes the common anchors again at every frame. In this section, we decode each such anchor once for a group of neighboring targets and share the resulting Gaussians, while every target keeps its own selection $S_t$. The obstacle is that decoding is view-dependent, so shared rows only approximate per-target decoding; we therefore share only within small groups of nearby poses. As illustrated in Figure~\ref{fig:cache}, \S\ref{subsec:Cache_Union} merges the requests of a group, \S\ref{subsec:Cache_Share} decodes them once under a common source pose, and \S\ref{subsec:Cache_Render} renders each target from the shared rows.

\subsection{Anchor Union}
\label{subsec:Cache_Union}
Since query locality holds between neighboring poses and all poses are known in advance, we partition the trajectory into consecutive, non-overlapping groups of configured length $K$. Each group
\begin{equation}
    E=\{\tau,\tau+1,\ldots,\tau+k-1\},
    \qquad 1\leq k\leq K,
    \label{eq:cache:group}
\end{equation}
contains $k$ targets; only the final group may be shorter than $K$. Within a group, we merge the selected sets by original anchor ID:
\begin{equation}
    U_E=\bigcup_{t\in E}S_t,
    \qquad m_E=|U_E|.
    \label{eq:cache:union}
\end{equation}
Each requested anchor appears once in $U_E$, so the $\sum_{t\in E}K_t$ per-target requests become $m_E$ decoding inputs. The \emph{reuse factor} $\rho_E=\sum_{t\in E}K_t/m_E$ ranges from $1$ (disjoint requests) to $k$ (identical requests), and it is exactly the factor by which \S\ref{subsec:Cache_Share} reduces decoding.
 
The union is a grouping query: grouping the request pairs $(i,t)$, $i\in S_t$, by anchor ID yields $U_E$ together with the \emph{membership table}
\begin{equation}
\label{eq:cache:demand_mask}
b_{i,t}=\begin{cases}1, & i\in S_t,\\ 0, & i\notin S_t,\end{cases}\qquad i\in U_E,\ t\in E.
\end{equation}
We evaluate it by sorting the pairs by anchor ID, which stores $U_E$ in deterministic ID order and returns the position of every request in $U_E$. This inverse mapping sets the bits of Eq.~\eqref{eq:cache:demand_mask} directly, in $O(\sum_{t}K_t\log\sum_{t}K_t)$ time and $k\,m_E$ bits. The union is only a decoding input: each target keeps its own $S_t\subseteq U_E$ as column $t$ of Eq.~\eqref{eq:cache:demand_mask}, and $U_E$ is never filtered by the frustum or LoD of a single pose, since that could drop anchors requested by another target.

\subsection{Gaussian Share}
\label{subsec:Cache_Share}
For a group $E$ and a source pose $\pi^{\mathrm{src}}_E$, the shared data is
\begin{equation}
\label{eq:cache:source_materialization}
\mathcal{G}^{\mathrm{src}}_E=\mathrm{Dec}_\theta\big(U_E,\pi^{\mathrm{src}}_E\big),\qquad N_{g,E}=|\mathcal{G}^{\mathrm{src}}_E|\le m_E n_o.
\end{equation}
Each row $r$ stores the attributes of Definition~\ref{def:3DGS} decoded under $\pi^{\mathrm{src}}_E$, its parent anchor ID $i_r$, and its candidate-offset ID $\kappa_r$; rows are read-only.
 
The decoder evaluates the $m_E\times n_o$ candidates and keeps those with positive opacity under $\pi^{\mathrm{src}}_E$. A prefix sum over this mask assigns each survivor its output position, and its array coordinates give $(i_r,\kappa_r)$, so the output is dense. A row reaches its targets through $i_r$, i.e., through row $(b_{i_r,t})_{t\in E}$ of Eq.~\eqref{eq:cache:demand_mask}, without storing its own copy.
 
\textbf{Source Pose.} Shared rows are exact only at the source pose and approximate the decode of every other target. If a target camera center lies within $\delta$ of the source one, the triangle inequality bounds the change in the viewing distance of an anchor at distance $D>\delta$ by $\delta$, and its viewing direction changes by at most $\arcsin(\delta/D)$. The deviation thus grows with $\delta$, so we minimize the largest $\delta$ in the group by placing $\pi^{\mathrm{src}}_E$ at the midpoint $\tau+(k-1)/2$: the central target for odd $k$, and for even $k$ a pose whose camera center is the midpoint of the two central ones (the decoder ignores orientation). If consecutive camera centers are at most $\eta$ apart, every target lies within $\eta(k-1)/2$ of the source, half the bound of decoding at the first target. The opacity mask, and hence the set of retained candidates, is also decided at the source. The resulting image distortion is evaluated against the quality requirement of Definition~\ref{def:Problem_Definition} in \S\ref{sec:experiments}.
 
\textbf{Shared Storage.} Decoding $U_E$ once reduces the group's neural work from $\Theta(n_o\sum_{t\in E}K_t)$ to $\Theta(n_o m_E)$, \ie by $\rho_E$. However, if each target then gathered its own copy of the rows it requests, the group would still copy $\Theta(\sum_{t\in E}|\mathcal{G}_t|)$ rows, up to $\Theta(kN_{g,E})$. We therefore keep a single copy in $\Theta(N_{g,E})$ space and let each target refer to it through its column of Eq.~\eqref{eq:cache:demand_mask}, as in the late materialization of column stores.

\subsection{Pose Render}
\label{subsec:Cache_Render}
For each target $t\in E$, its column of Eq.~\eqref{eq:cache:demand_mask} selects its logical input (Definition~\ref{def:Trajectory_Rendering}):
$\mathcal{G}_t=\big\{\, r\in\mathcal{G}^{\mathrm{src}}_E \;\big|\; b_{i_r,t}=1 \,\big\}$.
Since $b_{i_r,t}=1$ if and only if $i_r\in S_t$, and $U_E$ is never filtered by the source pose, $\mathcal{G}_t$ contains exactly the retained rows of the anchors in $S_t$. Sharing thus changes the attribute values a target consumes, but never the anchors it requested.
 
Projected positions, footprints, and depths, however, depend on $\pi_t$. Each target therefore projects the shared rows into target-private buffers, disables the rows outside $\mathcal{G}_t$ or invalid under $\pi_t$ (Section~\ref{subsec:Index_Basic}) by zeroing their radii and opacities, and completes S3 Image Rendering under $\pi_t$. A group thus projects $\Theta(kN_{g,E})$ rows rather than $\Theta(\sum_{t\in E}|\mathcal{G}_t|)$, which is the price of avoiding copies; a zero-radius row covers no image tile, so depth sorting and blending still process only $\mathcal{G}_t$. Since the shared rows are read-only and all target-specific results are private, the $k$ targets of a group can render concurrently, which Section~\ref{sec:Schedule} exploits.
 
\textbf{Group Length.} A longer group raises $\rho_E$ up to $k$, but it moves targets farther from the source, enlarges the shared data $N_{g,E}$ held until the last target of the group finishes, and delays decoding until all $k$ selections are ready. $K$ therefore balances decoding savings against fidelity and memory, and \S~\ref{sec:Schedule} bounds the shared data retained by concurrent groups.

\section{Decouple GPU--CPU Hybrid Pipeline}
\label{sec:Schedule}
This section exploits O3 of \S\ref{sec:Overview}, which concerns how this work is executed.  Figure~\ref{fig:Parallel} shows a further opportunity in how this work is executed. In the reference pipeline (Figure~\ref{fig:Parallel}(a)), \texttt{Selection}, \texttt{Materialization}, and \texttt{Rendering} are bound to each target on the GPU. Although the GPU can run several targets concurrently, materialization for a target still waits for its \texttt{Selection}, and the CPU stays idle because the projection-based \texttt{Selection} of existing methods runs on the GPU. Once \texttt{Selection} is expressed as the spatial queries of \S\ref{sec:Index}, it neither decodes Gaussian attributes nor depends on other targets' selections, so either processor can execute it. We exploit this separation to run \texttt{Selection} for the known trajectory on both CPU and GPU, then materialize and render groups from the collected results (Figure~\ref{fig:Parallel}(b)).

\subsection{Separating Selection from Rendering}
\label{subsec:Schedule_Execution}
We process the trajectory in \emph{batches} of complete groups of Eq.~\eqref{eq:cache:group}, and each batch in two stages separated by a barrier.

\textbf{Stage 1: Selection.} For every target $t$ of the batch, \texttt{Selection} computes the selected set $S_t$ of Definition~\ref{def:Trajectory_Rendering}. Each $S_t$ is an independent query result, so the CPU and the GPU both hold read-only copies of the anchor index $I_A$ and the occluder index $I_O$, execute the predicates of \S\ref{sec:Index}, and return IDs in the same increasing original-ID order. CPU workers use private traversal and output buffers, so queries for different targets run across CPU cores, while GPU queries run concurrently within the device's resources. Only the CPU-produced ID lists are transferred to the GPU, and no Gaussian attributes are decoded in this stage.

\textbf{Barrier.} Stage~2 starts only after every \texttt{Selection} of the batch has finished. A group needs the selections of all its targets to form $U_E$, and \S\ref{subsec:Schedule_Allocation} may assign the targets of one group to different processors; the barrier therefore keeps each processor's work independent of the other's progress. Its price is the storage of the selected IDs of the batch, $O(\sum_t K_t)$, which the batch size bounds.
 
\textbf{Stage 2: Materialization and Rendering.} For each group $E$, the GPU merges $(S_t)_{t\in E}$ into $U_E$ and $b_{i,t}$ (\S\ref{subsec:Cache_Union}), decodes $U_E$ once under the source pose (\S\ref{subsec:Cache_Share}), and lets each target render from the shared rows (\S\ref{subsec:Cache_Render}). Up to $B$ groups occupy GPU execution slots at a time, and each group renders up to $P_{\mathrm{rend}}$ targets concurrently. Groups enter the slots in trajectory order and are admitted only when their exact retained row count $N_{g,E}$ fits into the remaining shared-row budget $C$; the rows of a group are released after its last target finishes. Retained Gaussian data thus never exceed $C$, and since admission follows trajectory order, a later group never holds rows that an earlier group is waiting for.
 
\begin{figure}[t]
    \centering
    \includegraphics[width=1\columnwidth]{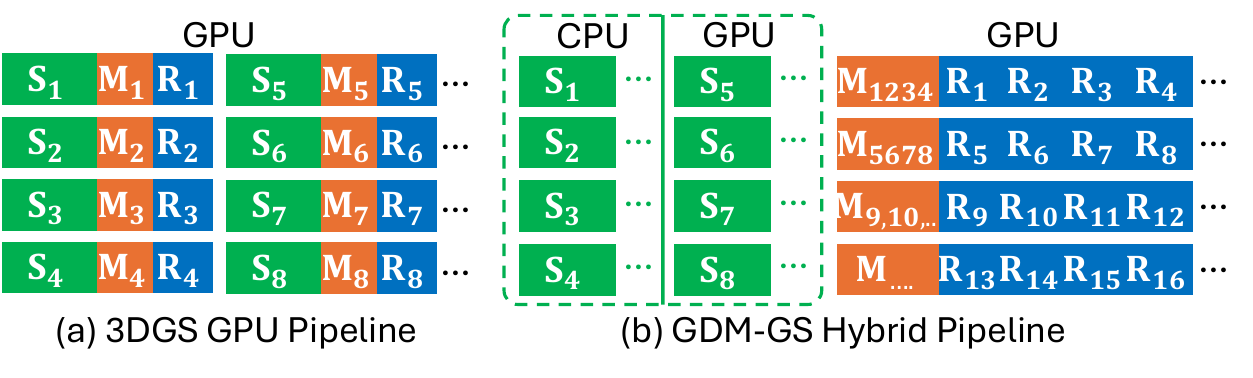}
    \caption{GPU-CPU Hybrid Parallelization}
    \label{fig:Parallel}
\end{figure}

The schedule changes neither $S_t$ nor $U_E$, so it introduces no approximation beyond the shared source pose of \S\ref{sec:Cache}, and the images do not depend on which processor computed each selection. Its savings complement those of \S\ref{sec:Cache}: Stage~1 shortens \texttt{Selection} by using both processors, as quantified in \S\ref{subsec:Schedule_Allocation}, while Stage~2 inherits the reduction of \texttt{Materialization} inputs from $\sum_{t\in E}K_t$ to $m_E$ per group; \texttt{Rendering} remains per target. Figure~\ref{fig:Parallel}(b) illustrates the resulting order: parallel CPU and GPU \texttt{Selection}s, one \texttt{Materialization} per group, and separate \texttt{Rendering}s for its targets.

 \subsection{Balancing CPU and GPU Selection}
\label{subsec:Schedule_Allocation}
Stage~1 ends when the slower processor finishes, so the batch should be split such that neither processor finishes substantially earlier than the other. Let $n$ be the number of targets in a batch, and let $x_C$ and $x_G$ be the numbers of \texttt{Selection} queries assigned to the CPU and the GPU, with $x_C+x_G=n$. We measure the effective \texttt{Selection} times per target, $c_C$ and $c_G$, at the chosen CPU worker count and GPU query concurrency, where $c_C$ includes producing and transferring the IDs. These are batch times divided by the number of queries rather than single-query latencies, so they account for parallel execution on each processor. For queries of comparable cost, the CPU and the GPU finish after about $c_Cx_C$ and $c_Gx_G$, so Stage~1 takes
\begin{equation}
\label{eq:schedule:allocation}
L=\max\big(c_Cx_C,\ c_Gx_G\big),
\end{equation}
which we minimize subject to $x_C+x_G=n$ and $x_C,x_G\ge 0$. Since the first term increases and the second decreases with $x_C$, the minimum balances the two sides:
\begin{equation}
\label{eq:schedule:solution}
x_C^*=\frac{nc_G}{c_C+c_G},\qquad
x_G^*=\frac{nc_C}{c_C+c_G},\qquad
L^*=\frac{nc_Cc_G}{c_C+c_G}.
\end{equation}
For integer counts, we take the better of $\lfloor x_C^*\rfloor$ and $\lceil x_C^*\rceil$ and assign the remaining queries to the GPU, which is optimal by the same monotonicity. Compared with GPU-only \texttt{Selection}, which takes $nc_G$, Stage~1 thus shrinks by the factor $1+c_G/c_C$, i.e., one plus the ratio of CPU to GPU throughput. For example, at 60 and 40 \texttt{Selection}s per second on the CPU and the GPU, a batch of 120 queries assigns 72 to the CPU and 48 to the GPU, and both finish after about 1.2\,s instead of 3\,s on the GPU alone.
 
Because of the barrier, this split shortens Stage~1 without affecting Stage~2: \texttt{Materialization} and \texttt{Rendering} start from the same sets $S_t$ whichever processor computed them, so their work is independent of the assignment. Eq.~\eqref{eq:schedule:allocation} therefore optimizes \texttt{Selection} rather than the complete trajectory latency. If poses differ substantially in query cost or ID volume, balanced counts need not balance the finish times. We therefore interleave the assignments along the trajectory, which spreads variation in view complexity without changing group membership, and re-measure the rates on representative poses for each batch.

%% file: sections/experiments.tex
\section{Experiments}
\label{sec:experiments}

\subsection{Experiment Settings}
\label{subsec:experiments_Setting}
\noindent\textbf{Scenes Data.}
As reported in Table~\ref{tab:exp:scenes}, we test on 8 BungeeNeRF scenes \cite{xiangli2022bungeenerf} at $1600\times900$. Each scence has approximately 1--1.3 million anchors, which will generate 10$\times$ more Gaussians $\#g_i$ because $n_o$ is set to 10. The last 3 columns report the relative rendering quality difference of GDM-GS compared with the SOTA Scaffold-GS \cite{lin2024scaffoldgs}, validating that the rendering quality is preserved.

\noindent\textbf{Query Trajectory.}
For each scene, we select 32-camera window and subdivide every adjacent camera interval into four steps, producing 125 target poses per scene and 1,000 poses in total, including 256 original photographic endpoints. Camera centers and focal lengths are linearly interpolated, while rotations use \textit{spherical linear interpolation} (SLERP)~\cite{shoemake1985}; the original endpoint poses are preserved. The windows were selected to represent high-reuse trajectory segments rather than random camera motion.

\begin{table}[t]
\centering
\caption{Scene Dataset Statistics ($n_o=10$)}
\label{tab:exp:scenes}
\vspace*{-4mm}
\resizebox{\columnwidth}{!}{%
\begin{tabular}{r|ccc|ccc}
\toprule
\textbf{Scene} &
\textbf{Avg \#$\pi$} &
\begin{tabular}[c]{@{}c@{}}\textbf{Anchors}\\ \textbf{$\times10^6$}\end{tabular} &
\begin{tabular}[c]{@{}c@{}}\textbf{\#$g_i$}\\ \textbf{$\times10^6$}\end{tabular} &
\begin{tabular}[c]{@{}c@{}}\textbf{$\Delta$PSNR}\\ \textbf{\% $\uparrow$}\end{tabular} &
\begin{tabular}[c]{@{}c@{}}\textbf{$\Delta$SSIM}\\ \textbf{\% $\uparrow$}\end{tabular} &
\begin{tabular}[c]{@{}c@{}}\textbf{$\Delta$LPIPS}\\ \textbf{\% $\downarrow$}\end{tabular}\\
\midrule
\textbf{Amsterdam AM} & 161 & 1.07 & 10.7 & -1.28 & -0.18 & +1.62 \\
\textbf{Barcelona BA} & 160 & 1.22 & 12.2 & +15.78 & +7.76 & -57.21 \\
\textbf{Bilbao BI}    & 129 & 1.01 & 10.1 & -1.11 & +0.84 & -7.09 \\
\textbf{Chicago CH}   & 160 & 1.11 & 11.1 & +0.23 & +0.67 & -5.84 \\
\textbf{Hollywood HO} & 125 & 1.06 & 10.6 & -0.50 & +0.51 & -8.95 \\
\textbf{Pompidou PO}  & 161 & 1.09 & 10.9 & -4.91 & -1.71 & +31.65 \\
\textbf{Quebec QU}    & 160 & 1.10 & 11.0 & +1.75 & +0.87 & -7.93 \\
\textbf{Rome RO}      & 158 & 1.29 & 12.9 & -0.24 & +0.15 & +4.14 \\
\midrule
\textbf{Average}      & 151.75 & 1.19 & 11.9 & \textbf{+1.21} & \textbf{+1.11} & \textbf{-6.2} \\
\bottomrule
\end{tabular}%
}
\scriptsize
$\Delta$PSNR, $\Delta$SSIM, and $\Delta$LPIPS are the relative changes of GDM-GS compared with Scaffold-GS \cite{lin2024scaffoldgs}.
\end{table}

\noindent\textbf{Environment.} 
All algorithms run on a server with an A100 80GB GPU and two Xeon Gold 6248R CPUs (48C/96T), using the runtime PyTorch~2.4/CUDA~12.4 and gsplat~1.4.0. %

\subsection{GDM-GS End-to-End Performance}
\label{sec:experiments:e2e}
\subsubsection{Rendering Quality}
We evaluate rendering quality using three complementary metrics: 1) Peak Signal-to-Noise Ratio (PSNR), which measures pixel-level reconstruction accuracy, with higher values indicating less error; 2) Structural Similarity (SSIM), which measures similarity in image structure, with values closer to 1 indicating greater similarity; and 3) Learned Perceptual Image Patch Similarity (LPIPS), which measures perceptual differences using learned visual features, with values closer to 0 indicating greater similarity. Table~\ref{tab:exp:scenes} reports the per-scene relative changes of GDM-GS compared with Scaffold-GS. Averaged across the eight scenes, PSNR and SSIM increase by 1.21\% and 1.11\%, respectively, while LPIPS decreases by 6.20\%. These average changes favor GDM-GS, although the effect varies across scenes. Barcelona shows substantial improvements, whereas Pompidou exhibits degradation in all three metrics. We further examine the effect of shared materialization on rendering fidelity in \S\ref{sec:experiments:materialization}.

\subsubsection{Rendering Efficiency}
As shown in Table \ref{tab:exp:e2e}, we report the compre the efficiency by rendering latency $ms/\pi$, \ie average time to render a pose. In general, GDM-GS is at least $2\times$ faster than the fastest baseline Octree-GS. In addition, we decompose the time of GDM-GS into three components: Selection, Materialization, and Rendering. It should be noted that the baselines' selection is coupled with their materialization and rendering, so it is impossible to decompose them into three parts and optimize them individually. Meanwhile, the rendering time of GDM-GS is the lowerbound of  rendering workload, because all the useless Gaussians are removed, and only the useful ones are rendered, so GDM-GS's rendering time is optimal and cannot be reduced. The materialization stage is already shared, and any more sharing would harm the quality, so its time is also optimal. As for selection, it is the only stage that can be further improved. Nevertheless, as the materialization and rendering are already optimal, even if selection time reduces to 0, the over speed up is still capped by 2.73$\times$, while our current speedup is already very close to this upperbound.

\begin{table}[t]
\centering
\caption{End-to-End Rendering Latency}
\label{tab:exp:e2e}
\vspace*{-4mm}
\resizebox{\columnwidth}{!}{%
\begin{tabular}{r|cc|cccc|c}
\toprule
&
\multicolumn{2}{c|}{\textbf{Baselines}} &
\multicolumn{4}{c|}{\textbf{GDM-GS}} &
\\
\cmidrule(lr){2-3}
\cmidrule(lr){4-7}
\textbf{Scene} &
\begin{tabular}[c]{@{}c@{}}\textbf{Scaffold}\\  ms/$\pi$\end{tabular} &
\begin{tabular}[c]{@{}c@{}}\textbf{Octree}\\  ms/$\pi$\end{tabular} &
\begin{tabular}[c]{@{}c@{}}\textbf{Sel}\\ ms/$\pi$\end{tabular} &
\begin{tabular}[c]{@{}c@{}}\textbf{Mat}\\ ms/$\pi$\end{tabular} &
\begin{tabular}[c]{@{}c@{}}\textbf{Ren}\\ ms/$\pi$\end{tabular} &
\begin{tabular}[c]{@{}c@{}}\textbf{Total}\\ ms/$\pi$\end{tabular} &
\begin{tabular}[c]{@{}c@{}}\textbf{Speedup}\\ \end{tabular} \\
\midrule
\textbf{AM} & 35.55 & 18.60 & 1.753 & 4.378 & 2.283 & \textbf{8.42} & \textbf{$2.21\times$} \\
\textbf{BA} & 11.06 & 11.21 & 1.411 & 2.283 & 1.863 & \textbf{5.56} & \textbf{$2.02\times$} \\
\textbf{BI} & 36.71 & 18.23 & 1.438 & 3.617 & 3.318 & \textbf{8.38} & \textbf{$2.18\times$} \\
\textbf{CH} & 46.62 & 18.16 & 1.659 & 3.439 & 3.419 & \textbf{8.52} & \textbf{$2.13\times$} \\
\textbf{HO} & 35.56 & 19.13 & 1.680 & 3.572 & 3.428 & \textbf{8.68} & \textbf{$2.20\times$} \\
\textbf{PO} & 31.34 & 16.26 & 1.402 & 4.223 & 1.576 & \textbf{7.21} & \textbf{$2.26\times$} \\
\textbf{QU} & 50.08 & 17.40 & 1.520 & 3.632 & 2.527 & \textbf{7.68} & \textbf{$2.27\times$} \\
\textbf{RO} & 55.58 & 18.75 & 1.700 & 3.608 & 3.211 & \textbf{8.53} & \textbf{$2.20\times$} \\
\midrule
\textbf{Avg}
& \textbf{37.81}
& \textbf{17.22}
& \textbf{1.516}
& \textbf{3.594}
& \textbf{2.703}
& \textbf{7.81}
& \textbf{$2.20\times$} \\
\midrule
\textbf{Opt}
& 37.81
& 17.22
& \textbf{0}
& \textbf{3.594}
& \textbf{2.703}
& \textbf{6.297}
& \textbf{2.73$\times$} \\
\bottomrule
\end{tabular}%
}
\end{table}

\subsection{Anchor Selection Performance}
\label{sec:experiments:selection}
\label{sec:experiments:bounds}
\label{sec:experiments:mesh}

\subsubsection{Anchor Selection Correctness}
Table \ref{tab:exp:bounds}'s $\#$ Missed Anchor shows how many anchors we miss per $\pi$ when we use only the original anchor as the smallest unit to build the index for anchor selection. If we time it by $n_o$, then millions of Gaussians are missed per view, which dramatically degrades rendering quality. In contrast, the proposed spatial bound-based method for anchor selection always has \#Missed Anchor = 0, which can also be shown to have no effect on final rendering quality.

\subsubsection{Anchor Selection Efficiency}
As shown in Table \ref{tab:exp:query-cost}, the Octree Selection is nearly the same as its end-to-end rendering time. This is because its Gaussian selection is tightly coupled with the rendering; out-of-sight Gaussians can only be excluded after materialization, and the occluded Gaussians can only be excluded after the rendering process. In contrast, our spatial frustum query determines which Gaussians are useful nearly $10\times$ faster, in 2ms, without materialization or rendering. It also proves the correctness and effectiveness of the decoupling selection.

\subsubsection{Occluder Pruning Effectiveness}
As shown in Table \ref{tab:exp:bounds}'s \#Pruned Anchor, the number of occluders varies by scenes and poses. It should be noted that all the tested scenes are general ones that do not have many objects blocking each other. Even in this general case, scenes like Amsterdam and Barcelona can still prune more than 30k anchors.

\subsubsection{Occluder Pruning Efficiency}
As shown in Table \ref{tab:exp:query-cost}, the Holed Frustum's query time brings the anchor selection time down to less than 1ms, which is more than 50\% lower than the complete frustum selection, and also proves the effectiveness of pruning the occluded areas during the anchor selection. However, retrieving the occluders takes more than 1ms, so the total selection time grows back to the original. Therefore, how to bring down the occluder retrieval time is a future direction worth digging into.

\begin{table}[t]
\centering
\caption{Anchor-Selection Error and Occluder Pruning}
\vspace*{-4mm}
\label{tab:exp:bounds}
\label{tab:exp:mesh-effect}
\resizebox{0.9\columnwidth}{!}{%
\begin{tabular}{c|cc||@{\hspace{10pt}}c|cc}
\toprule
\textbf{Scene} &
\begin{tabular}[c]{@{}c@{}}\textbf{\#Missed}\\ \textbf{Anchor}\end{tabular} &
\begin{tabular}[c]{@{}c@{}}\textbf{\#Pruned}\\ \textbf{Anchor}\end{tabular}  &
\textbf{Scene} &
\begin{tabular}[c]{@{}c@{}}\textbf{\#Missed}\\ \textbf{Anchor}\end{tabular}  &
\begin{tabular}[c]{@{}c@{}}\textbf{\#Pruned}\\ \textbf{Anchor}\end{tabular} \\
\midrule
\textbf{AM} & 203,330 & 369,432 &
\textbf{HO} & 236,191 & 4,057 \\
\textbf{BA} & 369,259 & 378,933 &
\textbf{PO} & 224,960 & 19,327 \\
\textbf{BI} & 244,374 & 64,206 &
\textbf{QU} & 175,376 & 56,293 \\
\textbf{CH} & 241,657 & 73,156 &
\textbf{RO} & 240,286 & 13,467 \\
\bottomrule
\end{tabular}%
}
\end{table}

\begin{table}[t]
\centering
\caption{Anchor Selection Efficiency (ms/$\pi$)}
\vspace*{-4mm}
\label{tab:exp:query-cost}
\resizebox{\columnwidth}{!}{%
\begin{tabular}{c|ccc|ccc}
\toprule
\textbf{Scene} &
\begin{tabular}[c]{@{}c@{}}\textbf{Octree}\\ \textbf{Selection }\end{tabular} &
\begin{tabular}[c]{@{}c@{}}\textbf{Frustum}\\ \textbf{Query}\end{tabular} &
\begin{tabular}[c]{@{}c@{}}\textbf{Speed}\\ \textbf{Up}\end{tabular}&
\begin{tabular}[c]{@{}c@{}}\textbf{Occluder}\\ \textbf{Retrieval}\end{tabular} &
\begin{tabular}[c]{@{}c@{}}\textbf{Holed}\\ \textbf{Frustum}\end{tabular} &
\begin{tabular}[c]{@{}c@{}}\textbf{Speed}\\ \textbf{Up}\end{tabular} \\
\midrule
\textbf{AM} & 18.102 & 2.145 & 8.44$\times$ & 1.312 & \textbf{0.916} & \textbf{$8.12\times$} \\
\textbf{BM} & 11.273 & 2.100 & 5.37$\times$ & 1.248 & \textbf{0.939} & \textbf{$5.15\times$} \\
\textbf{BI} & 19.529 & 2.003 & 9.74$\times$ & 1.264 & \textbf{0.828} & \textbf{$9.33\times$} \\
\textbf{CH} & 18.488 & 1.847 & 10$\times$ & 1.177 & \textbf{0.743} & \textbf{$9.63\times$} \\
\textbf{HO} & 19.633 & 2.360 & 8.32$\times$ & 1.427 & \textbf{1.082} & \textbf{$7.82\times$} \\
\textbf{PO} & 17.221 & 2.252 & 7.65$\times$ & 1.377 & \textbf{1.020} & \textbf{$7.18\times$} \\
\textbf{QU} & 17.545 & 1.924 & 9.12$\times$ & 1.278 & \textbf{0.742} & \textbf{$8.69\times$} \\
\textbf{RO} & 18.904 & 2.045 & 9.24$\times$ & 1.326 & \textbf{0.794} & \textbf{$8.92\times$} \\
\bottomrule
\end{tabular}%
}
\end{table}

\subsection{Materialization Performance}
\label{sec:experiments:materialization}
\label{sec:experiments:group-length}
\subsubsection{Influence of the Group Size $K$}
As shown in Table \ref{tab:exp:k}, for most of the scenes, as $K$ increases, the materialization time per $\pi$ also decreases, with the lowest taking less than 1/3 of the $K=1$ (single view)'s rendering time. This is because over 73\% of the anchors appear in every $\pi$ in the group, and only around 4\% are extra anchors unique to one $\pi$. Therefore, more than 73\% of the decoded Gaussians are shared across all four views, while only around 4\% of the workload comes from the extra add-ons. Together with those anchors only shared by 2 and 3 views, the actual materialization time increases from 1/4 of the total shared anchors to 1/3 of the total materialization time.

\subsubsection{Quality of the Shared Materialization}
As shown in the lower part of Table \ref{tab:exp:k}, we report the PSNR (the larger the better) of AM under $K=3, 4,$ and 6, using the first and the middle $\pi$ as the source pose. Specifically, as $K$ increases, PSNR decreases as expected: as more views share the anchors, the rendering quality worsens. In terms of source-view position, using the middle view consistently yields higher quality than using the first view, which matches expectations and intuition. Moreover, we also test the rendered results' quality against the groundtruth with the following criteria: direct PSNR $\geq40$\,dB, SSIM $\geq0.99$, LPIPS $\leq0.01$, depth relative MAE $\leq1\%$, and endpoint deterioration within $0.20$\,dB/$0.003$/$0.006$. It should be noted that only when all these strict conditions are met can a image pass the quality check. As shown in the last column of Table \ref{tab:exp:k}, among all the tested $k=4$ shared rendered images, only 1 of them fails the test. This guarantees the robust quality of our materialization.

\begin{table}[t]
\centering
\caption{Influence of Group Length $K$ and Source Pose}
\vspace*{-4mm}
\label{tab:exp:k}
\resizebox{\columnwidth}{!}{%
\begin{tabular}{c|ccc|cccc}
\toprule
\textbf{Scene}
& \begin{tabular}[c]{@{}c@{}}\textbf{$K$=1}\\ ms/$\pi$\end{tabular}
& \begin{tabular}[c]{@{}c@{}}\textbf{$K$=2}\\ ms/$\pi$\end{tabular}
& \begin{tabular}[c]{@{}c@{}}\textbf{$K$=3}\\ ms/$\pi$\end{tabular}
& \begin{tabular}[c]{@{}c@{}}\textbf{$K$=4}\\ ms/$\pi$\end{tabular}
& \begin{tabular}[c]{@{}c@{}}
    \textbf{Shared}\\
    \textbf{Anchor}
  \end{tabular}
& \begin{tabular}[c]{@{}c@{}}
    \textbf{Extra}\\
    \textbf{Anchor}
  \end{tabular}
& \begin{tabular}[c]{@{}c@{}}
    \textbf{Quality}\\
    \textbf{Pass}
  \end{tabular}
 \\
\midrule
\textbf{AM} & 12.555 & 8.298 & \textbf{3.926} & 4.378 & 73.45\%  & 3.65\% & 125/125\\
\textbf{BA} & 6.481 & 4.455 & \textbf{2.266} & 2.283 & 72.52\%  & 6.71\% & 125/125\\
\textbf{BI} & 11.047 & 7.051 & 4.151 & \textbf{3.617} & 73.36\%  & 4.23\% & 125/125\\
\textbf{CH} & 11.405 & 7.032 & 4.113 & \textbf{3.439} & 73.75\%  & 2.66\% & 125/125\\
\textbf{HO} & 11.985 & 7.541 & 4.363 & \textbf{3.572} & 73.18\%  & 4.69\% & 124/125\\
\textbf{PO} & 10.065 & 6.572 & \textbf{4.190} & 4.223 & 73.62\%  & 3.14\% & 125/125\\
\textbf{QU} & 11.005 & 8.200 & 4.163 & \textbf{3.632} & 73.36\%  & 4.40\% & 125/125\\
\textbf{RO} & 10.683 & 6.775 & 3.922 & \textbf{3.608} & 73.38\%  & 4.10\% & 125/125\\
\midrule\midrule
\multicolumn{8}{l}{\textbf{Amsterdam Source Pose Quality: Worst Direct PSNR (dB) $\uparrow$}} \\\hline
\multicolumn{2}{c|}{\textbf{Source Pose}}
& \multicolumn{2}{c}{\textbf{$K=3$}}
& \multicolumn{2}{c}{\textbf{$K=4$}}
& \multicolumn{2}{c}{\textbf{$K=6$}} \\
\midrule
\multicolumn{2}{c|}{\textbf{First} $\pi$}
& \multicolumn{2}{c}{42.62}
& \multicolumn{2}{c}{39.20}
& \multicolumn{2}{c}{35.15} \\
\multicolumn{2}{c|}{\textbf{Mid} $\pi$}
& \multicolumn{2}{c}{\textbf{48.80}}
& \multicolumn{2}{c}{\textbf{45.56}}
& \multicolumn{2}{c}{\textbf{41.56}} \\
\bottomrule
\end{tabular}%
}
\end{table}

\subsection{Scheduling Performance}
\label{sec:experiments:scheduling}
We evaluate the scheduling strategy in \S\ref{sec:Schedule} using 1,250 selection queries per scene. Table~\ref{tab:exp:scheduling} compares GPU-only, CPU-only, and balanced hybrid execution, where the CPU and GPU process disjoint subsets of the queries concurrently. Latency is measured as the total batch completion time divided by 1,250. The last two columns report the number of queries assigned to each processor

Although CPU-only execution is slower than GPU-only execution in every scene, hybrid execution consistently outperforms GPU-only execution. Across the eight scenes, average latency decreases from 1.906 to 1.383~ms per query, a 27.5\% reduction, corresponding to a $1.38\times$ speedup. Per-scene speedups range from $1.04\times$ on Pompidou to $2.47\times$ on Amsterdam. These results demonstrate that decoupling selection from Gaussian materialization allows CPU resources to contribute to trajectory processing and reduce selection completion time.

The allocation assigns 160 to 421 queries to the CPU and the remaining 829 to 1,090 queries to the GPU, reflecting their different processing rates. The CPU handles a smaller share so that its slower processing does not dominate batch completion. For example, Barcelona assigns 421 queries to the CPU, whereas Amsterdam assigns only 160. This unequal allocation allows the CPU to remove work from the GPU while limiting the waiting time before grouped materialization begins.

\begin{table}[t]
\centering
\caption{Selection Latency under Different Scheduling Configurations (ms/$\pi$).}
\label{tab:exp:scheduling}
\resizebox{0.9\columnwidth}{!}{%
\begin{tabular}{c|cc|ccc}
\toprule
\textbf{Scene} &
\begin{tabular}[c]{@{}c@{}}\textbf{GPU}\\ \textbf{Only}\end{tabular} &
\begin{tabular}[c]{@{}c@{}}\textbf{CPU}\\ \textbf{Only}\end{tabular} &
\begin{tabular}[c]{@{}c@{}}\textbf{Hybrid}\\ \textbf{Balanced}\end{tabular} &
\begin{tabular}[c]{@{}c@{}}\textbf{\#CPU}\\ \textbf{Queries}\end{tabular} &
\begin{tabular}[c]{@{}c@{}}\textbf{\#GPU}\\ \textbf{Queries}\end{tabular} \\
\midrule
\textbf{AM} & 3.514 & 10.900 & \textbf{1.423} & 160 & 1090 \\
\textbf{BA} & 1.674 & 3.401 & \textbf{1.108} & 421 & 829 \\
\textbf{BI} & 1.543 & 8.617 & \textbf{1.395 }& 200 & 1050 \\
\textbf{CH} & 1.629 & 6.630 & \textbf{1.195} & 160 & 1090 \\
\textbf{HO} & 1.819 & 8.530 & \textbf{1.449} & 200 & 1050 \\
\textbf{PO} & 1.820 & 12.783 & \textbf{1.426} & 200 & 1050 \\
\textbf{QU} & 1.631 & 10.590 & \textbf{1.369} & 160 & 1090 \\
\textbf{RO} & 1.619 & 12.143 & \textbf{1.372} & 160 & 1090 \\
\midrule
\textbf{Avg}. & 1.906 & 9.199 & \textbf{1.383} & 207.625 & 1042.375 \\
\bottomrule
\end{tabular}%
}
\end{table}

%% file: sections/relatedwork.tex
\section{Related Work}
\label{sec:related}

We organize NeRF and 3DGS methods along two non-exclusive axes: whether they change reconstruction state or rendering execution, and whether they optimize a single view or a camera trajectory.

\subsection{NeRF and Efficient NeRF}
\label{subsec:rw_nerf}
NeRF~\cite{mildenhall2020nerf} represents a scene as a continuous radiance field and renders each pose by evaluating the field along camera rays, repeating this work at every trajectory pose.
\textit{Reconstruction.}\enspace
Mip-NeRF~\cite{barron2021mipnerf} and Mip-NeRF 360~\cite{barron2022mip} improve anti-aliasing and scale handling, NeRF++~\cite{zhang2020nerfpp} separates foreground and background modeling for unbounded scenes, and Zip-NeRF~\cite{barron2023zipnerf} combines grid features with anti-aliasing. These methods improve the field queried by each view but retain pose-wise ray sampling. Block-NeRF~\cite{tancik2022blocknerf} and Mega-NeRF~\cite{turki2022meganerf} partition city-scale scenes into local fields for long camera paths.
\textit{Rendering.}\enspace
Neural Sparse Voxel Fields~\cite{liu2020neural}, PlenOctrees~\cite{yu2021plenoctree}, KiloNeRF~\cite{reiser2021kilonerf}, and Instant-NGP~\cite{muller2022instant} reduce the cost of each view through sparse voxels, octrees, small local networks, or hash encodings. Their spatial structures skip empty space along rays within one view, and trajectory serving still selects local fields and traces rays independently at each pose.

\subsection{3DGS and Efficient 3DGS}
\label{subsec:rw_3dgs}
3DGS~\cite{kerbl2023gs} represents a scene with explicit Gaussian primitives rendered by splatting~\cite{zwicker2002ewa}, which avoids ray marching but projects, sorts, and blends many primitives per frame. Scaffold-GS~\cite{lin2024scaffoldgs} and Octree-GS~\cite{ren2024octree} instead decode view-dependent Gaussians from anchors, the latter organized in a level-of-detail hierarchy. These representations distinguish persistent reconstruction units from pose-dependent emitted Gaussians.
\textit{Reconstruction.}\enspace
Pruning and compression~\cite{fan2024lightgaussian,niedermayr2024compressed,lee2024compact,ye2024compactgs,chen2024hac,chen2025hac++,fcgs2025,pcgs2025} shrink the stored Gaussian state, and hierarchical and virtualized representations~\cite{kerbl2024hierarchical3dgs,yang2025virtualized} vary its spatial resolution. LightGaussian removes low-impact primitives, while compact representations and learned compression encode retained attributes more efficiently. For large scenes and long paths, VastGaussian~\cite{lin2024vastgaussian}, CityGaussian~\cite{liu2024citygaussian}, and Horizon-GS~\cite{jiang2025horizongs} partition the scene into regions, while TideGS~\cite{tidegs2026} and A LoD of Gaussians~\cite{lodofgaussians2026} manage out-of-core storage and streaming, mainly for reconstruction scale. TideGS virtualizes visibility-sparse training state across storage tiers, whereas A LoD of Gaussians combines hierarchical selection with cache-aware streaming.
\textit{Rendering.}\enspace
gsplat~\cite{ye2025gsplat}, FlashGS~\cite{feng2025flashgs}, and hardware rasterization~\cite{yuan2025hardware} accelerate per-view rasterization, while OccluGaussian~\cite{liu2025occlugaussian} and Proxy-GS~\cite{proxygs2026} cull invisible regions or occluded anchors for each view. OccluGaussian uses camera co-visibility to partition scenes, whereas Proxy-GS uses proxy depth for anchor culling and training-time densification. For trajectories, GS-Cache~\cite{gs-cache-2025} combines caching, scheduling, and optimized kernels, and LODGE~\cite{kulhanek2025lodge} reuses precomputed region-level active sets across adjacent views. LODGE combines these active sets with level-of-detail selection and chunk-based rendering to reduce repeated selection overhead.

\textit{Positioning.}\enspace
In both families, what a view needs is decided by per-view computation over rays, primitives, or anchors, and hierarchies mainly serve level-of-detail selection; to the best of our knowledge, none answers selection as a range query over an index of conservative bounds or removes occluded space from the query before decoding. Trajectory systems reuse cached data or precomputed working sets, whereas GDM-GS shares the view-dependent decoding of anchors across nearby poses and pipelines CPU selection with GPU rendering.

\subsection{Spatial Data Management}
\label{subsec:rw_db}
Our techniques build on classical database methods. Spatial access methods such as R-trees~\cite{guttman1984rtree} and region octrees~\cite{samet1984quadtree} answer range queries over bounded objects by pruning whole groups; we apply them to anchor bounds and to solid space, where the query region is a frustum with holes. Sharing decoding across the targets of a group resembles multiple-query optimization~\cite{sellis1988mqo}, which evaluates the common work of concurrent queries once, and sharing rows through membership columns follows late materialization in column stores~\cite{abadi2006materialization}. CPU--GPU co-processing has been studied for relational operators~\cite{he2008gpujoins}; we instead pipeline index traversal on the CPU with neural decoding and rendering on the GPU.

%% file: sections/conclusion.tex
\section{Conclusion}
\label{sec:Conclusion}
In this paper, we studied efficient trajectory rendering for structured 3DGS and proposed GDM-GS, a spatial data management framework. By reformulating anchor selection as spatial queries, GDM-GS decouples data retrieval from Gaussian materialization and rendering. This separation enables spatial indexes to filter unnecessary inputs before decoding, neighboring views to share materialized data, and CPUs and GPUs to jointly process selection queries. GDM-GS thus optimizes a trajectory as a collection of related requests, using their spatial relationships and shared inputs to reduce both the work required and the time to execute it.